\newcommand{\diracslash}[1]{#1\llap{/\kern2pt}}

\newcommand{\be}{\begin{equation}}
\newcommand{\ee}{\end{equation}}
\newcommand{\bea}{\begin{eqnarray}}
\newcommand{\eea}{\end{eqnarray}}
\newcommand{\ba}[1]{\begin{array}{#1}}
\newcommand{\ea}{\end{array}}

\documentclass[prd,aps,floats,nofootinbib,tightenlines,showpacs]{revtex4}
\usepackage{epsfig,graphicx,pstricks}
\usepackage{psfrag}
\usepackage{color}
\usepackage{amsmath}
\usepackage{amsfonts}
\usepackage{amssymb}
\usepackage{color, soul}
\usepackage{textcomp}
\usepackage{multirow}
\usepackage{natbib}
\usepackage{subfigure}
\begin{document}
\title{Net-Charge Fluctuations in Finite Volume PNJL Model: A Probe for the QCD Critical Point}
\author{A.~Sarkar}
\email{amal.sarkar@cern.ch}
\author{P.~Deb}
\email{paramita.dab83@gmail.com}
\author{Bidhan Mandal}
\email{mandal.bidhan1440@gmail.com}
\author{R.~Varma}
\email{raghava.varma@cern.ch}
\affiliation{*School of Physical Science, Indian Institute of Technology Mandi, Kamand, Mandi - 175005, India}
\affiliation{{\dag\S}Department of Physics, Indian Institute of Technology Bombay, Powai, Mumbai- 400076, India}
\affiliation{{\ddag}Department of Physics, Ramakrishna Mission Residential College (Autonomous), Narendrapur, Kolkata - 700103, India}

\begin{abstract}
The QCD Critical Point is a pivotal feature of the phase diagram of strongly interacting matter. Signatures of the critical point are expected to manifest through the non-monotonic behavior of higher-order moments of conserved quantities, such as net-baryon ($\Delta B$), net-charge ($\Delta Q$), and net-strangeness ($\Delta S$), as a function of collision energy. These moments are connected to the thermodynamic susceptibilities, as well as to the correlation length developed in the system, which diverges at the critical point. The non-monotonic behavior of higher-order moments and their volume-independent products near the critical region
supports the presence of a critical point in a finite system existing for a finite time, due to their sensitivity to critical fluctuations. These fluctuations are believed to provide key evidence in the search for the QCD Critical Point. We present the higher order moments, such as mean (M), variance $(\sigma^2)$, skewness (S), and kurtosis $(\kappa)$ and their volume-independent moment products $(M/\sigma^{2}, s\sigma, \kappa\sigma^{2})$ of net-charge multiplicity distributions in the three-flavor finite volume, finite density Polyakov loop enhanced Nambu-Jona-Lasinio (PNJL) model. The work has been performed at energies similar to RHIC BES energies from 7.7 GeV to 200 GeV, including 2.4 and 3 GeV in the present model. Our findings are compared with net-charge data from the Solenoidal Tracker at RHIC (STAR) across various collision energies to explore signals of the QCD critical point. Additionally, we contrast our results with predictions from the Ultra-relativistic Quantum Molecular Dynamics (UrQMD) model, the Hadron Resonance Gas (HRG) model, and available lattice QCD data. The present results offer a useful tool for extracting the freeze-out parameters in the heavy-ion collision by comparing them with the STAR net-charge result and other net-charge theoretical models. 

\end{abstract}
\maketitle

\section{\label{sec:level1}Introduction
}
Quantum Chromo Dynamics (QCD) delineates a strong interaction of quarks and gluons, which evinces a rich phase structure of strongly interacting hot and dense nuclear matter at finite temperatures and densities. The QCD Critical Point represents a unique landmark in the phase diagram of Quantum Chromodynamics (QCD), marking the boundary between the crossover region and the first-order phase transition. It signifies the point where the nature of the phase transition in strongly interacting matter changes, characterized by large fluctuations. The search for this point is of great interest in high-energy physics, as its discovery would provide fundamental insights into the behavior of nuclear matter under extreme temperature and density conditions, offering a deeper understanding of the early universe and neutron stars. The principle aim of the Beam Energy Scan (BES) program at the BNL Relativistic Heavy Ion Collider (RHIC) is to search for the existence of the QCD critical point, which is also known as the second-order phase transition point. This critical point, as has been suggested, exists in non-vanishing baryon chemical potential in the $T$-$\mu_B$ phase diagram of QCD \cite{1}. Due to the asymptotic features of QCD, it is anticipated that at very high temperatures and density, the handron will dismantle itself to liberate quarks and gluons. This deconfinement phase, where relevant degrees of freedom are quarks and gluons, is known as quark-gluon plasma (QGP) \cite{2}, and it has been found at RHIC. Lattice QCD calculations have revealed that at high temperatures and zero baryon chemical potential $(\mu_B=0)$ a simple crossover transition from QGP to hadronic matter occurs \cite{3, 4, 5, 6}. However, various QCD based models suggest that at large $\mu_B$ and low temperatures, the quark-hadron phase transition is of first order \cite{7, 8}. Therefore, it is expected that the first-order phase transition line between hadronic to QGP matter in the QCD phase diagram should end at a point at finite $\mu_B$ and finite temperature where the phase transition is second order, known as the critical point (CP) or critical end point (CEP). Experimentally finding the critical point, which is very close to the freeze-out point, would be an important landmark for the study of the phase structure of QCD matter \cite{9, 10}. The Beam Energy Scan program at RHIC has been ongoing since 2010 \cite{11}. The STAR collaboration has reported the centrality and energy dependence of the moment of net-charge multiplicity distribution for Au+Au collision at $\sqrt{S_{NN}}=7.7, 11.5, 19.6, 27.0, 39.0, 62.4$, and $200$ GeV \cite{12}. In this program, the chemical freeze-out temperature and baryon chemical potential have been tuned by varying the RHIC colliding energy $(\sqrt{S_{NN}})$ from 200 GeV to 7.7 GeV, which covers the $\mu_B$ from 410 MeV to 20 MeV \cite{13}. Suitable experimental observables of conserved quantities that are expected to be sensitive to the proximity of the freeze-out point.\\

The Higher order moments (Mean (M), variance ($\sigma^2$), skewness (S), and kurtosis ($\kappa$)) and their volume independent moment products ($s\sigma$, $\kappa \sigma^2$) of the conserved quantities, such as net-baryon $(\Delta B)$, net-charge $(\Delta Q)$, net-strangeness $(\Delta S)$ number are thought to be useful observables for searching the signature of the phase transition and critical point. The measured multiplicity distribution is used to calculate the statistical moment in terms of statistics. Let N denote any entry in the data sample. The average of N over the whole event ensemble is symbolized by $\left\langle N\right\rangle $. The fluctuation of N around its mean net-particle multiplicity is represented by $\delta N$=N-$\left\langle N\right\rangle$. The higher order moments describe the event-by-event distributions of conserved quantities within a limited acceptance, are given as follows: Mean M=$\left\langle N\right\rangle$, variance 
$\sigma^2$ = $\left\langle(\delta N)^2\right\rangle$, skewness S = $\left\langle(\delta N)^3\right\rangle$/$\sigma^3$ and kurtosis $\kappa$ = $\left\langle(\delta N)^4\right\rangle$/$\sigma^4$ - 3. The $r^{th}$ central moment is defined as:
	\begin{equation}\label{1}
	\mu_{r}=\left\langle \left( \delta N\right)^r \right\rangle	
	\end{equation}
The cumulants of a given event by event particle multiplicity distribution for each centrality can be expressed in terms of moments as follows
\begin{equation}\label{2}
    \begin{split}
	C_1=\mu=\left\langle N\right\rangle\\
	C_2=\mu_2 \\
	C_3=\mu_3\\
	C_4=\mu_4-3\mu_2^{2}
    \end{split}
	\end{equation}
A recursion relation summaries the expression $C_n(n>3)$ as \cite{49}	
	\begin{equation}\label{3}
	C_n=\mu_n-\sum_{m=2}^{n-2}\binom{n-1}{m-1}C_m\mu_{\left( n-m\right)}
	\end{equation}
various moments discussed in connection to cumulants are as  
\begin{equation}\label{4}
	M=C_1\text{ ; }\sigma^2=C_2\text{ ; }S=\frac{C_3}{(C_2)^{3/2}}\text{ ; }\kappa=\frac{C_4}{(C_2)^{2}}
    \end{equation}
The properties of the net-particle multiplicity distribution are defined by the moments. An expectation operator of the multiplicity density is described by the first-order moment. The second-order moment, also known as variance, indicates the susceptibility of the measured particle multiplicity distribution. Skewness is the term used to describe the normalization of the third-order moment. Skewness provides information regarding the direction of the data set's variation. The asymmetry of the distribution is described by skewness. The terms "kurtosis" refer to the normalization of the fourth-order moment. Kurtosis compares the peaked distributions with normal distributions.  Analytical calculations are used to compute statistically significant moments from the definition of the partition function. The moments of net conserved quantum quantities are related to the thermodynamic susceptibilities, $\chi_i^{\left( n\right) }$, where \emph{n} denotes the order of the susceptibility and \emph{i} represents the type of conserved quantum number. They are also related to the high power of the correlation length of the system. For net-charge distribution, \emph{i} becomes the charge quantum number (\emph{Q}). These susceptibilities are expressed in terms of cumulants ($C_n$) as
	 \begin{equation}\label{5}
	 \chi_i^{\left( n\right) } = \frac{1}{VT^3}C_n
	 \end{equation}	
where V is the volume of the cubic system and T is the temperature. When susceptibilities are related to moments, a volume term that depends on the size of the system appears. Therefore, it is difficult to compare several systems and collision centralities. The susceptibility ratios, moment products, or ratios of cumulants of net-particle multiplicity distributions are calculated to cancel out the volume term. They are written as
	 \begin{equation}\label{6}
	 	\frac{\sigma^2}{M}=\frac{\chi_i^{\left( 2\right) }}{\chi_i^{\left( 1\right) }}\\
	 	S\sigma=\frac{\chi_i^{\left( 3\right) }}{\chi_i^{\left( 2\right) }}\\
	 	\kappa\sigma^2=\frac{\chi_i^{\left( 4\right) }}{\chi_i^{\left( 2\right) }}
	 \end{equation}
The calculation of the product of moments using the ratios of various-order cumulants is given as follows
	 \begin{equation}\label{7}
	 \frac{\sigma^2}{M}=\frac{C_2}{C_1}\text{ ; }S\sigma=\frac{C_3}{C_2}\text{ ; }\kappa\sigma^2=\frac{C_4}{C_2}\text{ ; }\frac{\kappa\sigma}{S}=\frac{C_4}{C_3}
	 \end{equation}\\

The non-monotonic behaviour in the fluctuation of the conserved quantities $\Delta B$, $\Delta Q$, and $\Delta S$ number, as a function of colliding energy, could be one of the characteristic signatures about the location of the CP \cite{14, 15, 16, 17, 18}. The most important observables being studied in the continuing low energy runs at RHIC are higher-order moments of the event-by-event distribution of conserved quantities like Mean $(M)$, Standard deviation $(\sigma)$, Skewness $(S)$, and Kurtosis $(\kappa)$. These moments of conserved charge distribution have a close relation to the higher order thermodynamic susceptibilities and higher power of the correlation length $(\xi)$ produced by the system $(\sigma^2 \sim \xi^2, S \sim \xi^{4.5}, \kappa \sim \xi^{7})$ \cite{19}. The higher-order susceptibilities and the correlation length of the system seem to diverge at CP within the thermodynamic limit. So the order parameter exhibits extraordinarily large fluctuations on the long wavelength with critical slowdown i.e., an increase in relaxation time due to the finite time and finite size of the fireball created in heavy ion collision. Due to the finite size effect and critical slowdown, correlation length $(\xi)$ rises to a maximum value in the range of 1.5 to 3 fm \cite{20, 21}. Because of independence from the undetermined interaction volume, moment products $(M/\sigma^{2}, s\sigma, \kappa\sigma^{2})$ for different orders have been constructed for comparison of the theoretical calculation with experiments. The finite volume of the QCD matter created in the heavy ion collision experiments depends on the size of the colliding matter, the center of mass energy, and collision centrality. A large number of attempts have been made by different centers of mass energy $(\sqrt{S_{NN}})$ and centrality measurements of HBT radii to determine the finite volume of the system \cite{22}. These findings imply that freeze-out volume increases with centrality. The effects of finite freeze-out volume have been discussed by several QCD-based theoretical models, such as the non-interacting bag model \cite{23}, chiral perturbation theory \cite{24}, the linear sigma model \cite{25, 26}, and the Nambu-Jona-Lasinio model \cite{27, 28, 29}. A few first-principles studies of pure gluon theory on a space-time lattice were executed to study the possibility of significant finite-size effects \cite{30, 31}. Recently, the induction of a charged pion condensation phenomenon in dense baryonic matter was performed due to finite volume effects in the 1+1 dimension NJL model. There has been a recent study regarding the thermodynamic properties of strongly interacting matter in a finite volume using the Polyakov loop extended Nambu-Jona-Lasinio (PNJL) model \cite{32, 33, 34}. This model was developed from the NJL model by introducing a background gauge field (Polyakov loop) with the NJL Lagrangian. Deconfinement and chiral symmetry breaking, two of QCD's primary distinctive features, are encapsulated in the PNJL model. In the PNJL model, it has been shown that the temperature at the critical point at zero baryon density decreases as the volume of the system decreases. So CEP is shifted towards the higher $\mu_B$ and lower temperature domains at low volumes \cite{35}. Due to divergence behaviour at CP, the increasing fluctuations of the higher-order moments serve as more reliable indicators of the phase transition from hadronic to partonic states. Various QCD-inspired models, like the Polyakov loop coupled quark-meson (PQM) model \cite{36}, the renormalized group developed version, the 2 flavors PNJL model with three momentum cutoff regularization \cite{37, 38}, the higher flavors PNJL model \cite{32, 34, 35} with or without finite volume effects, along with simplified lattice QCD \cite{39} have been built to calculate fluctuations with respect to quark chemical potential. Recently, a re-parametrized three flavor PNJl model \cite{40} based on realistic continuum limit calculation \cite{41, 42, 43, 44} for lattice QCD showed satisfactory quantative results for second and fourth order susceptibilities of baryon number with lattice data at finite temperature and zero density region. However, it is not possible to investigate the whole QCD phase diagram due to the limitation of lattice data at finite density.\\
The present study will emphasize the three flavors finite volume finite density Polyakov loop enhanced Nambu-Jona-Lasinio (PNJL) model with six quark interaction to explore the fluctuations of the higher moments and their volume independent moment products of net-charge with energies similar to RHIC BES energies along with 2.4 and 3 GeV energies. Lattice QCD calculations inspire to extraction of the freeze out parameter, so it is expected that higher order moment and their volume independent moment products of net-charge will exhibit fluctuations as they are much more sensitive to the presence of CP of the QCD phase diagram. The non monotonic behavior of the crucial fluctuations would therefore provide strong support for CP. 

In this paper, we investigate the fluctuations of higher-order cumulants $C_1$, $C_2$, $C_3$, and $C_4$—which are directly related to moments (mean ($M$), variance ($\sigma^2$), skewness ($S$), and kurtosis ($\kappa$))—and their volume-independent cumulant ratios (moment products ($s\sigma$, $\kappa\sigma^2$)) of net-charge distributions within the three-flavor finite volume, finite density PNJL model. The study examines six-quark interactions with system sizes of R = 2 fm and 4 fm, while eight-quark interactions are also explored for these system sizes. Studies on heavy-ion collisions suggest that larger fireballs with longer lifespans are produced at higher collision energies compared to lower-energy collisions, with system sizes potentially reaching up to 10 fm in heavy-ion collisions. To facilitate comparison with experimental heavy-ion collision data, three finite-volume systems with radii of R = 2 fm, and 4 fm are considered, enabling a comparison between finite-volume systems across different sizes. 
The present work is organized as follows:
Section~\ref{sec:level2} summarizes the mathematical definitions for the observables used in data analysis. Section~\ref{sec:level3} briefly outlines the formalism of the PNJL model. In Section~\ref{sec:level4}, we present our results of the fluctuations of higher order moment and their volume independent moment products obtained from the PNJL model for searching critical point. Finally, the conclusion is given in the Section~\ref{sec:level5}.
The study is conducted at energies similar to the RHIC Beam Energy Scan (BES) to compare with experimental findings. The results are compared with STAR experimental data on net-charge \cite{45} fluctuations, as well as with predictions from the HRG \cite{46} HIJING \cite{47}, and UrQMD \cite{48} models. The behavior of the data is analyzed, and the fluctuations as a function of energy provide insights into the experimental expectations for the critical region.

\section{\label{sec:level2}THE PNJL MODEL}
The Polyakov-Nambu-Jona-Lasinio (PNJL) model is one of the QCD motivated successful phenomenological models that helps in combining the confinement and chiral symmetry breaking properties in a single framework. The physics of spontaneous chiral symmetry breaking can be created using a four quark interaction term in the NJL Lagrangian. Because of the dynamical generation of fermion mass, chiral symmetry breaks spontaneously. The physics of color confinement, however, is not adequately described in the NJL model as the gluons have been integrated out. In the PNJL model the chiral point coupling between quarks, present in the NJL Lagrangian, and a temporal gluon field that symbolizes Polyakov loop dynamics are accounted for to explain the phenomenon of gluon physics. The Polyakov line is given by 
\begin{equation}\label{8}
L(\vec{x})={p}\exp[i\int^{\beta}_{_0}d\tau A_{_4}(\vec{x},\tau)]
\end{equation}
where the temporal component of Euclidean gauge field $(\vec{A},A_{_4})$ is $A_{_4} = iA_{_0}$, in which strong coupling constant $g_{_s}$ has been absorbed, $p$ indicates path ordering and $\beta=1/T$ with Boltzmann constant $K_{_B}=1$. The Polyakov line $L(\vec{x})$ converts to a Polyakov loop field with unitary charge under global $Z(3)$ symmetry. The Polyakov loop field is then specified by $\Phi=(Tr_{_c}L)/N_{_c}$ and its Hermitian conjugate $\bar{\Phi}=(Tr_{_c}L^{\dag})/N_{_c}$. The effective theory that underlies the Polyakov loops represents gluon dynamics. Therefore, the Polyakov loop potential can be demonstrated as 
\begin{equation}\label{9}
 \frac{\mathcal{U}^{\prime}(\Phi,\bar{\Phi},T)}{T^{4}}=\frac{\mathcal{U}(\Phi,\bar{\Phi},T)}{T^{4}}-\kappa ln[ \jmath(\Phi,\bar{\Phi})]
\end{equation}
where $\mathcal{U}(\Phi)$ is the Landau-Ginzburg type potential commensurate with the $Z(3)$ global symmetry, expressed as \cite{50}
\begin{equation}\label{10}
 \frac{\mathcal{U}(\Phi,\bar{\Phi},T)}{T^{4}}=-\frac{b_{_2}(T)}{2}\bar{\Phi},\Phi-\frac{b_{_3}}{6}(\Phi^{3}+\bar{\Phi}^{3})
 +\frac{b_{_4}}{4}(\bar{\Phi}\Phi)^{2}
\end{equation}
where $ b_{_2}(T)=a_{_0}+a_{_1}(\frac{T_{_0}}{T})+a_{_2}(\frac{T_{_0}}{T})^{2}+a_{_3}(\frac{T_{_0}}{T})^{3}$ and $T_{_0}$ is the critical temperature for deconfinement phase transitions according to pure gauge lattice theory.
$b_{_3}$ and $b_{_4}$ being constant. The term $\frac{\mathcal{U}(\Phi,\bar{\Phi},T)}{T^{4}}$ is the vandermonde term which replicates the effect of $SU(3)$ Haar measure and is denoted by,
\begin{equation}\label{11}
  \jmath[\Phi,\bar{\Phi}]=[(\frac{27}{24\pi^{2}})(1-6\Phi\bar{\Phi}+4(\Phi^{3}+\bar{\Phi}^{3})-3(\Phi\bar{\Phi})^{2}]
\end{equation}
$\jmath[\Phi,\bar{\Phi}]$ is also known as the Vandermonde determinant. This determinant is explicitly independent of space-time. The value of the dimensionless parameter $\kappa$ will be determined phenomenologically and generally will have a dependence on some temperature and/or chemical potential. The corresponding parameters were fitted with the help of some physical quantities as a function of temperature from the Lattice QCD calculations of pure gauge theory \cite{50}. The set of chosen values for six quark interactions are, $T_0=175$ MeV, $a_0=6.75, a_1=-9.0, a_2=0.25, b_3=0.805, b_4=7.555$ and $\kappa=0.1$ \cite{40}. In the NJL model, the interaction between quarks is accounted for with the exception of the substituent of a covariant derivative that includes the temporal background gauge field. Thus, the Lagrangian of the PNJL model for 2+1 flavors is given by
\begin{equation}\label{12}
\begin{split}
  \mathcal{L}=\sum_{_{f=u,d,s}}\bar{\psi}_{_f}\gamma_{_\mu}iD^{\mu}\psi_{_f}-\sum_{_f} m_{_f}\bar{\psi}_{_f}\psi_{_f}
  +\sum_{_f}\mu_{_f}\gamma_{_0}\bar{\psi}_{_f}\psi_{_f}+\\
  \frac{g_{_s}}{2}\sum_{_{a=0,...,8}}[(\bar{\psi}\lambda^{a}\psi)^{2}
  + (\bar{\psi}i\gamma_{_5}\lambda^{a}\psi)^{2}]-g_{_D}[det\bar{\psi}_{_f}P_{_L}\psi_{_{f^{\prime}}}\\
  + det\bar{\psi}_{_f}P_{_R}\psi_{_{f^{\prime}}}]-\mathcal{U}^{\prime}[\Phi,\bar{\Phi},T]
\end{split}
\end{equation}
where u or d or s, respectively, are the flavors symbolized by $f$. $g_{_s}$ and $g_{_D}$ are four quark and six quark interaction coupling constants respectively and $g_{_D}$ is set to zero for two flavors. The diagonal components of the current quark mass matrix are denoted by $m_{_f}$, with $m_{_f}=diag(m_{_u},m_{_d},m_{_s})$. The left-handed and right-handed chiral projectors are respectively described by the matrices $P_{_{L,R}}=(1\pm\gamma_{_5})/2$. $D^{\mu}=\partial^{\mu}-iA^{\mu}$ indicates the covariant derivative. The terms have their usual meaning, described in detail in Refs. \cite{51, 52, 53, 32, 35}. Here, the strong interaction coupling is absorbed in the gauge field $A^{\mu}$ and is given by $A^{\mu}=\delta^{\mu}_{_0}A_{_0}$ (Polyakov gauge); in Euclidean notation $A_{_0}=-iA_{_4}$. $\lambda^{a}$ represents the flavor $SU_{_f}(3)$ Gell-Mann matrices $(a=0,1,..,8)$, where $\lambda^{0}=\surd(\frac{2}{3})I$. There are similarities between the theory part of the NJL model and the BCS theory of superconductors. BCS theory suggests that the electrons act as pair coupled by lattice vibration and condense into a state which creates a gap in the energy spectrum. Analogous to this, in the NJL model, a composite operator $\langle\bar{\psi}_{_f}\psi_{_f}\rangle$ that yields a nonzero vacuum expectation value leads to quark condensation because of the dynamical breaking of $SU(3)_{_L}\times SU(3)_{_R}$ symmetry to $SU(3)_{_V}$ with a chiral limit.
The quark condensate is given by
\begin{equation}\label{13}
 \langle\bar{\psi}_{_f}\psi_{_f}\rangle=-N_{_c}\mathcal{L} t_{_{y\rightarrow x^{+}}}(trS_{_f}(x-y)),
\end{equation}
where the trace is over color and spin states. Effective quark masses can be easily calculated from Lagrangian using the mean field method. From the so-called gap equation, these masses are derived.
\begin{equation}\label{14}
  M_{_f}=m_{_f}-g_{_s}\sigma_{_f}+g_{_D}\sigma_{_{f+1}}\sigma_{_{f+2}}
\end{equation}
where $\sigma_{_f}=\langle\bar{\psi}_{_f}\psi_{_f}\rangle$ represents the chiral condensate of the quark with flavor $f$. Here, it is considered $\sigma_{_f}=\sigma_{_u}$, then $\sigma_{_{f+1}}=\sigma_{_d}$ and $\sigma_{_{f+2}}=\sigma_{_s}$. Similarly if $\sigma_{_f}=\sigma_{_d}$, then $\sigma_{_{f+1}}=\sigma_{_s}$ and $\sigma_{_{f+2}}=\sigma_{_u}$, if $\sigma_{_f}=\sigma_{_s}$, then $\sigma_{_{f+1}}=\sigma_{_u}$ and $\sigma_{_{f+2}}=\sigma_{_d}$. When the temperature and chemical potential are both to zero, the expression for $\sigma_{_f}$ is expressed as \cite{54}
\begin{equation}\label{15}
\sigma_{_f}=-\frac{3M_{_f}}{\pi^{2}}\int^{\Lambda}_{_0}\frac{p^{2}}{\sqrt{(p^{2}+M^{2}_{_f})}}dp
\end{equation}
where the three-momentum cut-off is $\Lambda$. Due to a finite dimensional coupling constant, this model is therefore not renormalizable. So, the cut-off regulator has been used to regulate the model. Here, the mean field approximation (MFA) is used to obtain $\phi$, $\bar{\phi}$ and $\sigma_{_f}$ fields. Dynamically breaking chiral symmetry results in the appearance of $N^{2}_{_f}-1$ Goldstone bosons. These Goldstone bosons are pions and kaons which have been detected experimentally. Their masses and decay width are utilized to fix the NJL model parameters. The parameters for six quark interaction where the eight quark coupling constants $g_1$ and $g_2$ are taken to zero, has been incorporated to the present work. Additionally, it has been demonstrated that the lowest four-quark interaction term in the quark sector spontaneously breaks chiral symmetry to create a stable vacuum. However, the six quark interaction term is invariant under $SU(3)_{_L}\times SU(3)_{_R}$. But the $U(1)_{A}$ symmetry is broken by this interaction term. So the six quark interaction term imitate the QCD chiral anamoly and destabilize the vacuum \cite{55}. The parameters used in the present work for the stability of the PNJl model are given as follows: $m_u=5.5$ MeV, $m_s=134.76$ MeV, $\Lambda=631$ MeV, $g_s \Lambda^2=3.67$, $g_D \Lambda^5=9.33$, $g_1 \times 10^{-21} (MeV^{-8})=0$, $g_2 \times 10^{-22} (MeV^{-8})=0$. A non-zero low momentum cut off $p_{min}$=$\pi$/R=$\lambda$ has been chosen for implementation of the PNJL model in the finite volume where the lateral size of the finite cubic volume system is R. Nonetheless, it is necessary to take into account the effects of surface and curvature. The proper boundary conditions (periodic for bosons and anti periodic for fermions) would lead to a infinite sum over discrete momentum values. But in order to simplify, the infinite sum has been regarded as an integration over continuous variation of momentum even though with a lower momentum cut-off and the surface and curvature effects have been disregarded. No modifications should be made to the mean field parameters due to finite size effect. The philosophy for this study is to clutch the known physics at zero temperature, zero chemical potential and infinite volume fixed. It is explained that volume (V) can be considered as thermodynamic variable in the same footing as T and $\mu$. Any variation resulting from a change in either of these thermodynamic parameters was converted into a change in the effective fields of $\sigma_f$, $\phi$ etc and, through them, to the meson spectra. It is naturally expected that the model parameters, which are set by the values of meson masses and decay constants, would strictly follow at T=0, $\mu$=0, and V=$\infty$. This would result in no change to the Polyakov loop potential as well as the meanfield part of the NJL model. So, it could be preferable to continue utilizing the saddle point approximation in the PNJL model to analyse finite volumes. To study the thermodynamic properties of QCD, the thermodynamic potential for the multi-fermion interaction in the mean field (MFA) approximation of the PNJL model is given below \cite{54}.
\begin{equation}\label{16}
\begin{split}
  \Omega^{\prime}(\Phi,\bar{\Phi},M_{_f},T,\mu_{_f})=\mathcal{U}^{\prime}[\Phi,\bar{\Phi},T]+
  2g_{_s}\sum_{{f=u,d,s}}(\sigma^{2}_{_f}-\frac{g_{_D}}{2}\sigma_{_u}\sigma_{_d}\sigma_{_s})
  \\-T\sum_{_n}\int\frac{d^{3}p}{(2\pi)^{3}}Tr ln\frac{S^{-1}(i\omega_{_n},\vec{p})}{T}
\end{split}
\end{equation}
Matsubara frequencies $\omega_{_n}=\pi T(2n+1)$ for fermions are used here. In momentum space, the expression for the inverse quark propagator can be written as
\begin{equation}\label{17}
  S^{-1}=\gamma_{_0}(p^{0}+\hat{\mu}-iA_{_4})-\vec{\gamma}.\vec{p}-\hat{M}
\end{equation}
Using the identify $Tr ln(X) = ln det(X)$, we get
\begin{equation}\label{18}
\begin{split}
\Omega^{\prime}=\mathcal{U}^{\prime}[\Phi,\bar{\Phi},T]+
2g_{_s}\sum_{{f=u,d,s}}(\sigma^{2}_{_f}-\frac{g_{_D}}{2}\sigma_{_u}\sigma_{_d}\sigma_{_s})
-6\sum_{_f}\int^{\Lambda}_{_\lambda}\frac{d^{3}p}{(2\pi)^{3}}E_{_{pf}}\Theta(\Lambda-|\vec{p}|)-2\\
\sum_{_f}T\int^{\infty}_{_\lambda}\frac{d^{3}p}{(2\pi)^{3}}ln[1+3(\Phi+\bar{\Phi}exp(\frac{-(E_{_{pf}}-\mu_{_f})}{T}))
exp(\frac{-(E_{_{pf}}-\mu_{_f})}{T})+exp(\frac{-3(E_{_{pf}}-\mu_{_f})}{T})]-2\\
\sum_{_f}T\int^{\infty}_{_\lambda}\frac{d^{3}p}{(2\pi)^{3}}ln[1+3(\bar{\Phi}+\Phi exp(\frac{-(E_{_{pf}}+\mu_{_f})}{T}))
exp(\frac{-(E_{_{pf}}+\mu_{_f})}{T})+exp(\frac{-3(E_{_{pf}}+\mu_{_f})}{T})]\\
\Omega^{\prime}=\Omega-\kappa T^{4}ln\jmath[\Phi,\bar{\Phi}]
\end{split}
\end{equation}
where $ E_{_{pf}}=\sqrt{(p^{2}+M^{2}_{_f})}$ is the energy single quasiparticle. $\sigma_f$=$\langle\bar{\psi}_{_f}\psi_{_f}\rangle$ represents the chiral condensate of the quark with flavor f with $\sigma_f^2$=($\sigma_u^2+\sigma_d^2+\sigma_s^2$). In the above integral, the vacuum integral has a finite limit $\Lambda$ where as the integral having dependence on the medium are extended upto $\infty$. The lower cut off $\lambda$=$\pi$/R serves as a restriction for lower limit of both vacuum integral and medium dependent integral. Two sets of finite volume systems have been implemented in the PNJL model for comparison with infinite volume systems. The original model was created using an infinite volume system. However, as the fireball produced by the heavy-ion collision has a finite volume, a finite volume system with the radii R=2fm and R=4fm can be taken into consideration to better compare with the experimental data. The interaction terms in the Lagrangian of the three-flavor NJL model combine four and six quarks. Therefore, two parameter set (2fm6q and 4fm6q) of net-charge are considered in the present study for searching critical point.\\

The center of mass collision energy $(\sqrt{S_{NN}})$, that regulates the temperature and density of the produced fireball, is an important parameter for analyzing the observables in the experiments. In order to obtain collision energy dependence, a parameterization of the freeze-out conditions between $\sqrt{S_{NN}}$ and other thermodynamic variables is required. Although different methods for new parameterization of the freeze-out condition as a function of collision energy in the T-$\mu_B$ plane have been offered. In order to not be limited to the T-$\mu_B$ plane alone, the old method has been utilized, which makes use of the values of baryon, charge, and strangeness chemical potentials that are available with respect to freeze-out temperature and BES energy. The old method guaranteed evaluation of all chemical potential values. The following parameterization has been used in the current study to determine the freeze-out curve T($\mu_B$) in the T-$\mu_B$ plane and the colliding energy dependency baryon chemical potential ($\mu_B$) in a nucleus-nucleus collision \cite{13}.
\begin{equation}\label{19}
\begin{split}
 T(\mu_B) = a-b\mu_B^{2}-c\mu_B^{4}\\
 \mu_B(\sqrt{S_{NN}}) = d/(1+e\sqrt{S_{NN}})
 \end{split}
\end{equation}
where a = (0.166$\pm$0.002) $GeV$, b = (0.139$\pm$0.016) $GeV^{-1}$, c = (0.053$\pm$0.021) $GeV^{-3}$, d = (1.308$\pm$0.028) $GeV$ and e = (0.273$\pm$0.008) $GeV^{-1}$. 

\section{\label{sec:level3}RESULTS}
In this study, we present the results of higher-order cumulants, $C_1$, $C_2$, $C_3$, and $C_4$ (moments, $M$, $\sigma$, $S$, $\kappa$) of net-charge and their volume independent ratios $C_1/C_2$, $C_3/C_2$, and $C_4/C_2$ (products, $M/\sigma^2$, $s\sigma$, $\kappa\sigma^{2}$) as a function of center of mass energy in the formalism of the three-flavor finite volume, finite density Polyakov loop extended Nambu-Jona-Lasinio (PNJL) model with six quark interactions. These moments of the conserved charges have a connection with the susceptibilities, which are significantly influenced by the experimentally formed correlation length $(\xi)$ in the heavy ion collision. The correlation length is expected to diverge at CP in the thermodynamic limit; therefore, the susceptibilities and moments also diverge at the critical point. There is a correlation between the critical point and the magnitude of the non-Gaussian fluctuation of the conserved quantities. So, these fluctuations of conserved values as a function of colliding energy exhibit a non-monotonic behaviour, which can be considered a signature of CEP. Due to their connection with susceptibilities, these higher moments are sensitive to the system volume. Therefore, moment products (cumulant ratios) can be treated as system volume independent observables. The presented results for various cumulants and their volume independent rations from the PNJL model have been obtained at a fixed quark chemical potential for different energies similar to the RHIC Beam Energy Scan (BES) energies (7.7, 11.5, 14.5, 19.6, 27.0, 39.0, 62.4, 130.0, and 200 GeV) along with 2.4 and 3 GeV, respectively. 

\begin{figure*}[htb]
  \centering
   [$\mathsf{(a)}$]{{\includegraphics[width=8.0cm]{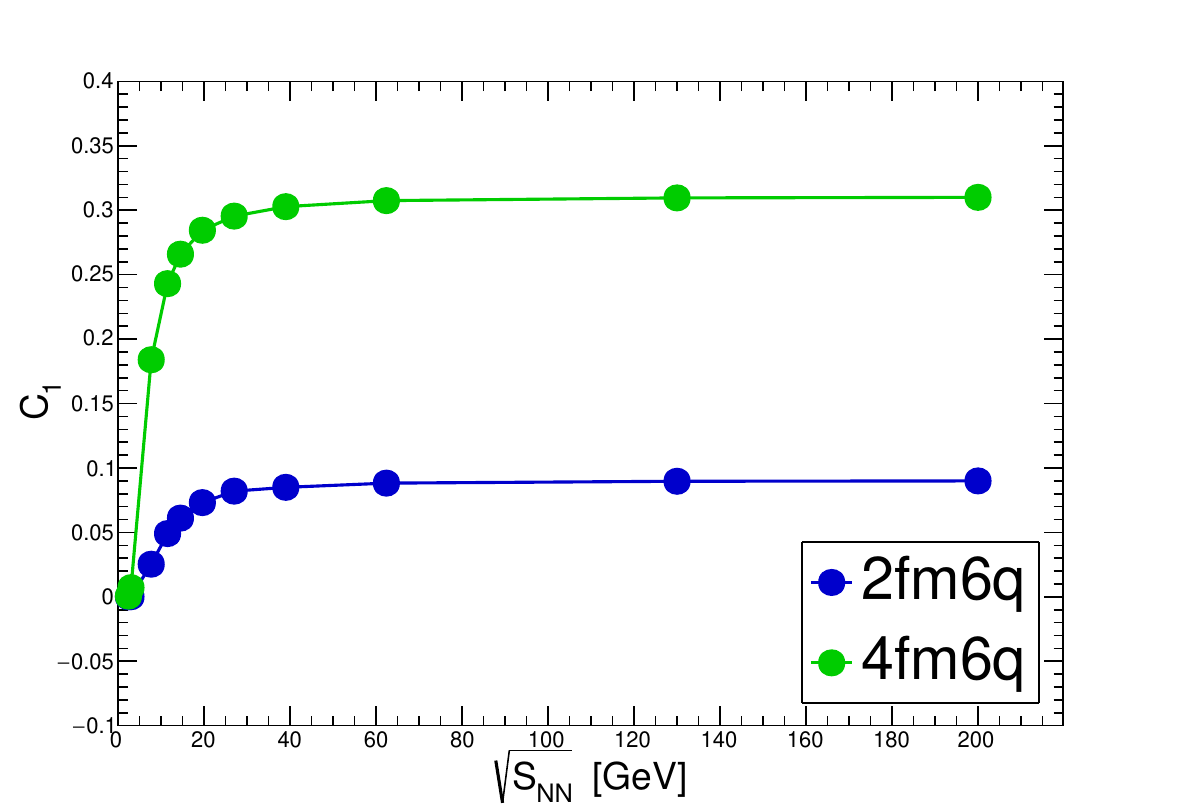} }}
   [$\mathsf{(b)}$]{{\includegraphics[width=8.0cm]{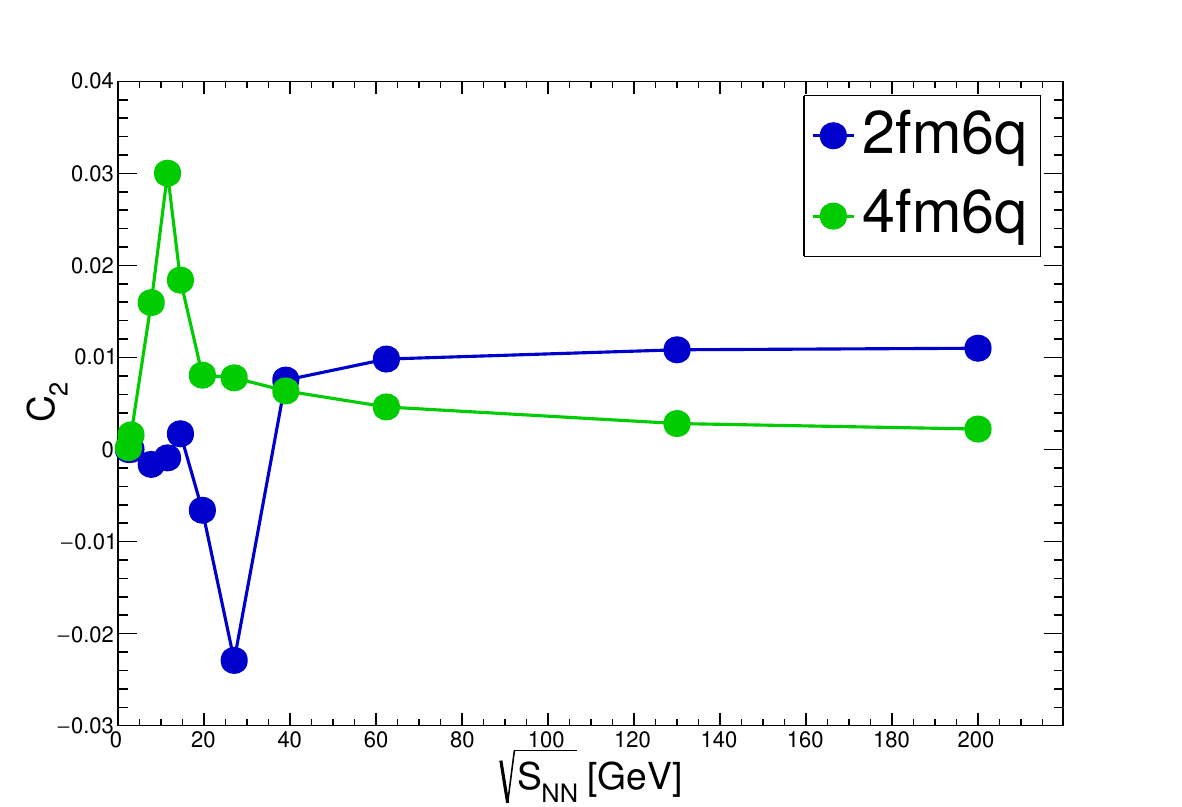} }}
   [$\mathsf{(c)}$]{{\includegraphics[width=8.0cm]{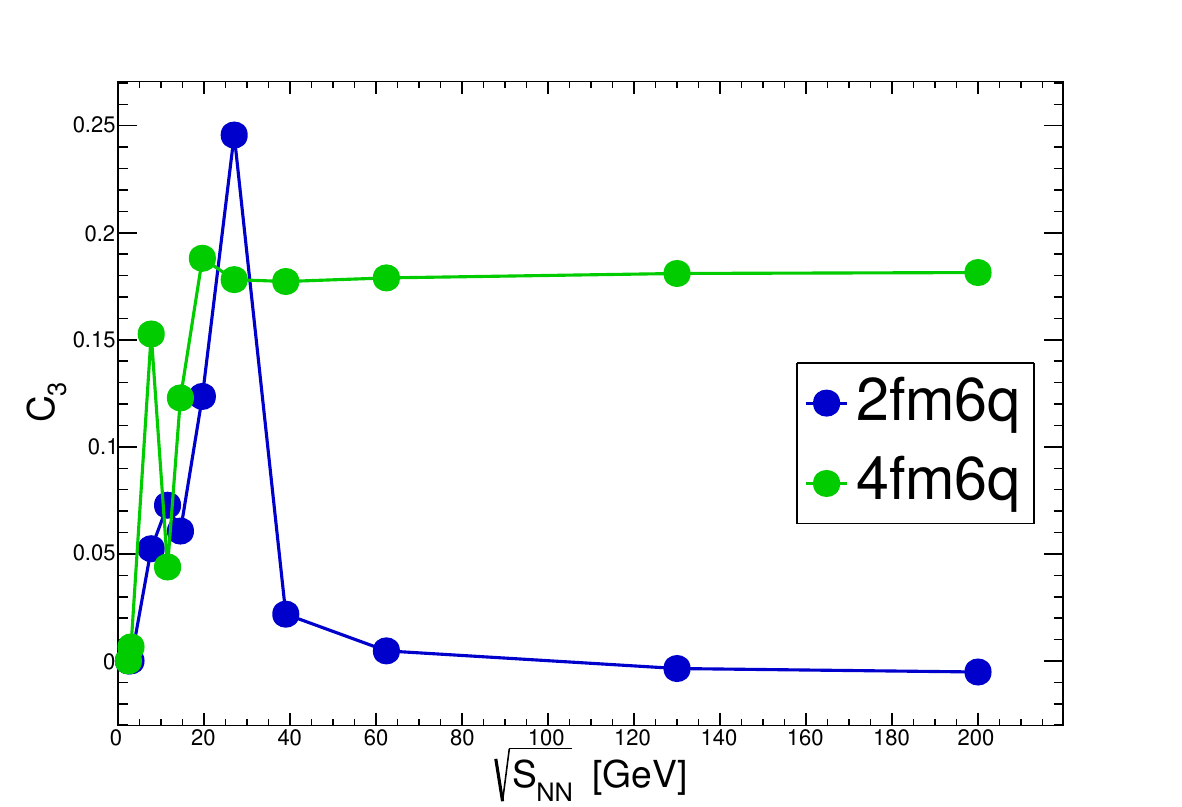} }}
   [$\mathsf{(d)}$]{{\includegraphics[width=8.0cm]{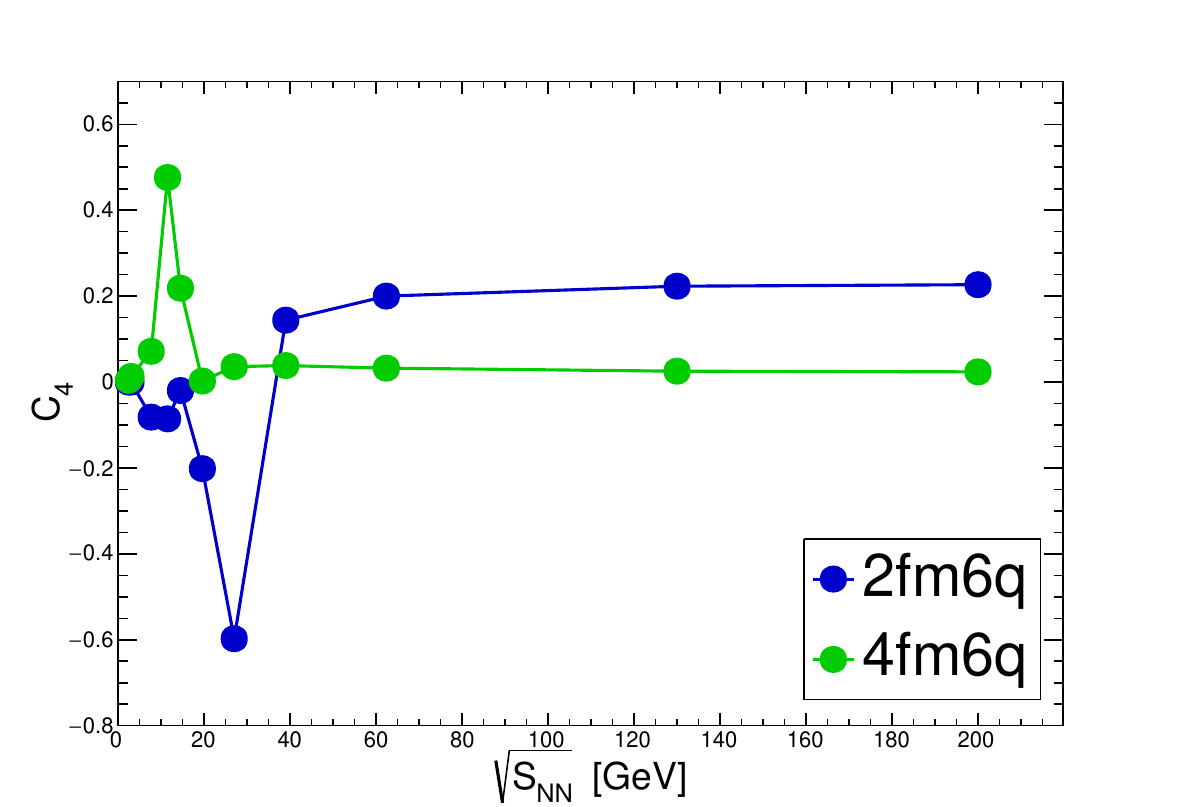} }}
    \caption{(color online) (a) First-order cumulant ($C_{1})$ (b) Second-order cumulant ($C_{2}$)  (c) Third-order cumulant ($C_{3}$) (d) Fourth-order cumulant ($C_{4}$) as a function of energy in PNJL model with 6 quarks interactions for infinite volume systems with lateral size R = 2fm and R = 4fm. The system with 6 quark interactions with R = 2fm is shown in blue and R = 4fm is shown in green color. }
    \label{fig:1}
\end{figure*}

Figures~1(a)--1(d) show the first-to-fourth-order cumulants ($C_{1}, C_{2}, C_{3}, C_{4}$) of net-charge fluctuations computed using the PNJL model at various collision energies (in GeV). The results are presented for $\sqrt{s_{NN}} = 2.4$, 3, 7.7, 11.5, 14.5, 19.6, 27.0, 39.0, 62.4, 130.0, and 200~GeV. Figure~1(a) displays the first-order cumulant ($C_1$) as a function of energy. It is observed that the difference between the 2~fm 6-quark (blue) and 4~fm 6-quark (green) results increases at lower energies, with the 4~fm curve rising more rapidly. At higher energies, both curves become largely energy-independent. Figure~1(b) presents the second-order cumulant ($C_2$). Here, the 2~fm and 4~fm systems with the same interaction type exhibit opposite behaviors: the 4~fm data shows a peak near 11.5~GeV, while the 2~fm data exhibit a dip around 27~GeV. Beyond 39~GeV, both curves level off, becoming nearly constant, with the 2~fm values exceeding those of the 4~fm system. In Figure~1(c), the third-order cumulant ($C_3$) is shown. The 4~fm curve displays mild fluctuation in the low-energy region (below 20~GeV), while the 2~fm curve features a prominent peak around 27~GeV. At higher energies, both curves run nearly parallel and appear independent of energy. Figure~1(d) shows the fourth-order cumulant ($C_4$) as a function of energy. While the 2~fm and 4~fm results follow qualitatively similar trends, their magnitudes differ. The behavior of $C_4$ with energy is broadly consistent with that of $C_2$, although the differences between the 2~fm and 4~fm results are more pronounced.

\begin{figure*}[htb]
  \centering
   [$\mathsf{(a)}$]{{\includegraphics[width=8.2cm]{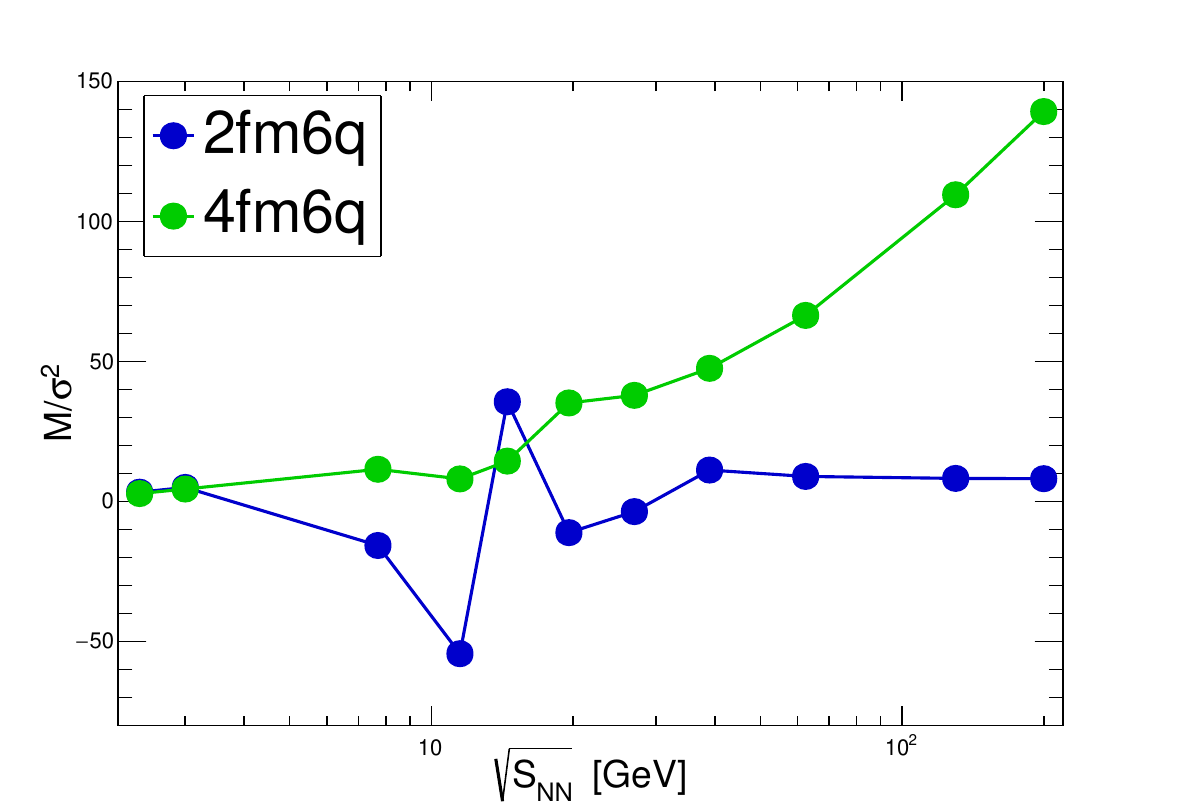} }}
   [$\mathsf{(b)}$]{{\includegraphics[width=8.2cm]{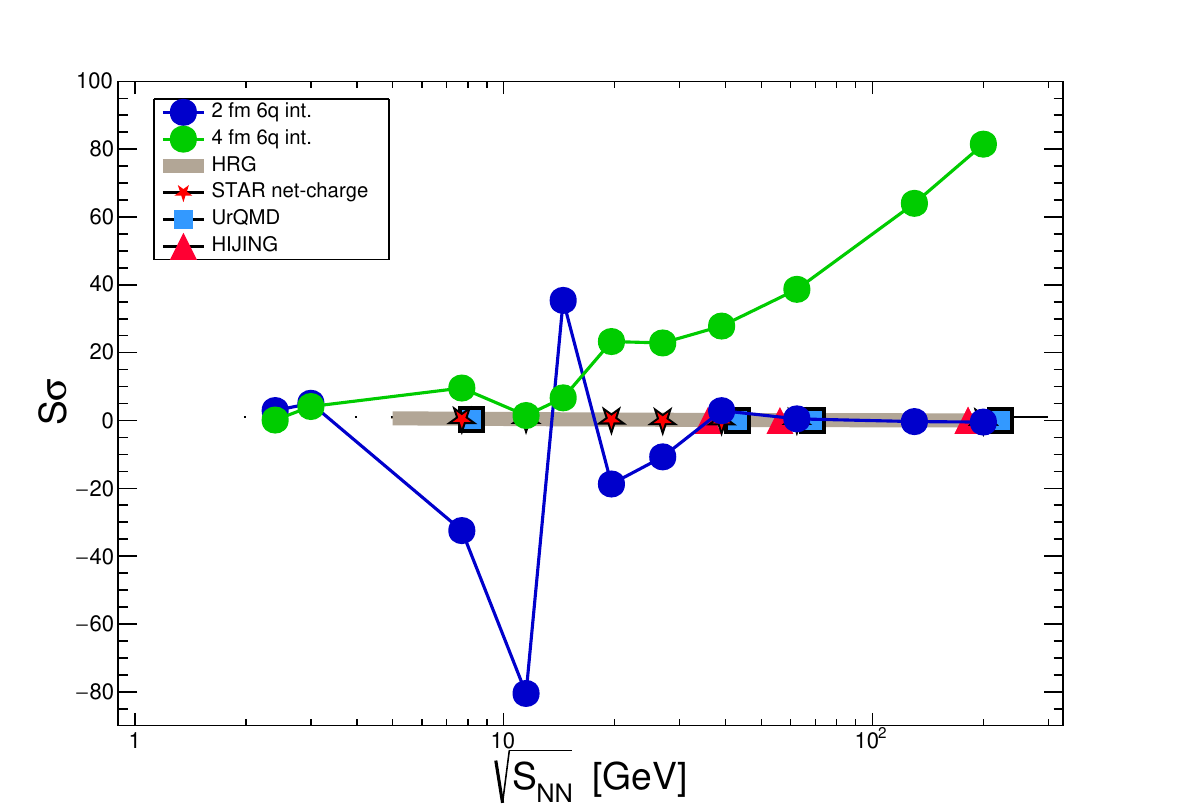} }}
    \caption{(color online) Beam-energy dependence of moment product (a)$M/\sigma^{2}$($\frac{C_{1}}{C_{2}}$)(b)$S\sigma$($\frac{C_{3}}{C_{2}}$) in the PNJL model with 6 quark interactions for finite volume systems with radius R=2fm and 4fm. For the system with 6q-PNJL model, blue color curve represents 2fm and green color curve represents 4fm results. $S\sigma$ as a function of beam energy are compared with the net-charge data of STAR experiments in red with black star. The results are compared with HRG, UrQMD and HIJING model calculations in grey, azure with black square and pink respectively}
    \label{fig:2}
\end{figure*}

Figure~2 presents the beam-energy dependence of two moment products: (a) $M/\sigma^{2}$ ($C_{1}/C_{2}$) and (b) $S\sigma$ ($C_{3}/C_{2}$), calculated using the PNJL model with 6-quark interactions for finite volume systems with radii $R=2$~fm and $R=4$~fm. In panel~2(a), the $M/\sigma^{2}$ values for the 2~fm system exhibit strong fluctuations at lower energies (below 30~GeV), while the 4~fm results increase gradually with energy beyond this point. The 2~fm results, however, remain largely energy-independent in the high-energy regime. Panel~2(b) shows the $S\sigma$ moment product. The 2~fm data exhibits pronounced non-monotonic fluctuations below 40~GeV, but becomes energy-independent above this threshold. As the system size increases from 2~fm to 4~fm, these fluctuations diminish, and the 4~fm data begins to show a monotonic increase in $S\sigma$ beyond 40~GeV. For comparison, results from HRG, STAR, UrQMD, and HIJING models for $S\sigma$ and $\kappa\sigma^{2}$ (shown in Figure~3) have been included using data from Refs.~\cite{45, 46, 47, 48}. The comparison reveals that the 4~fm data deviates significantly from the STAR net-charge experimental results and other model predictions. In contrast, the 2~fm results qualitatively follow the trends of experimental and model data at higher energies, although with quantitative differences. However, at lower energies, the 2~fm data fails to capture the experimental and theoretical behavior observed in STAR and model studies.


\begin{figure*}[htb]
  \centering
   [$\mathsf{(a)}$]{{\includegraphics[width=8.2cm]{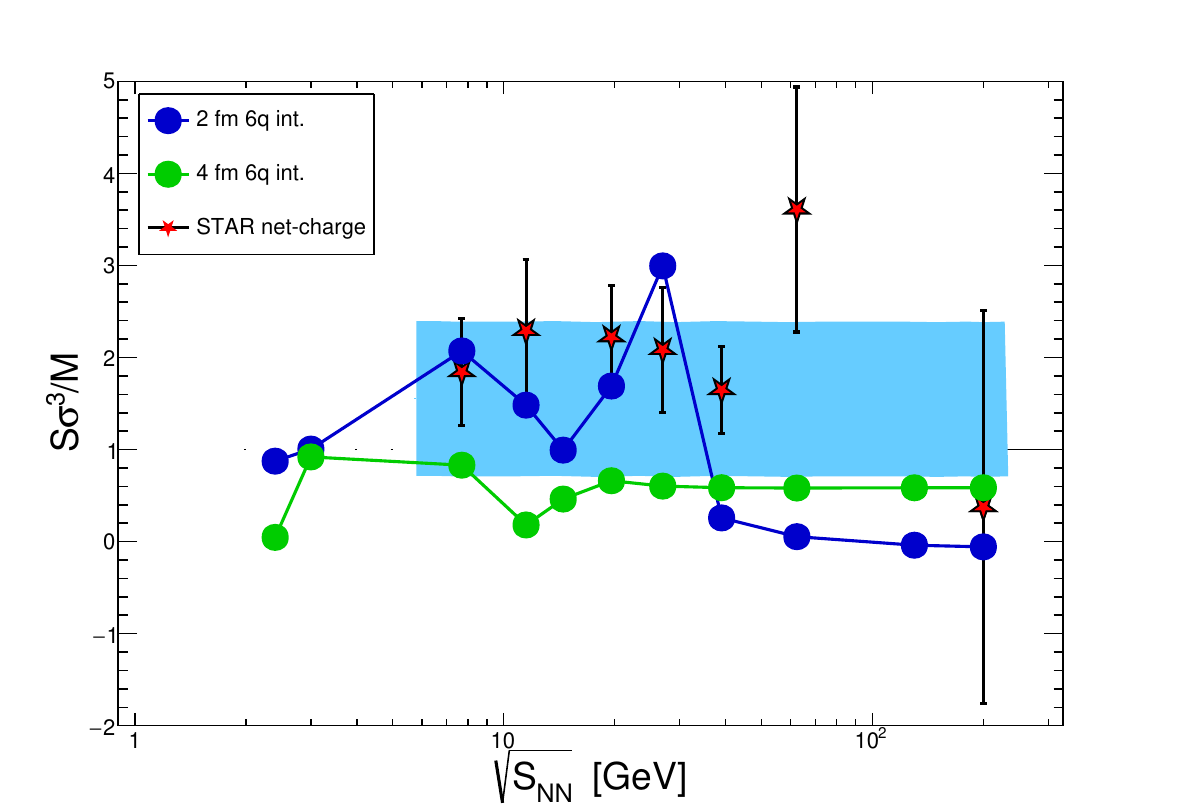} }}
   [$\mathsf{(b)}$]{{\includegraphics[width=8.2cm]{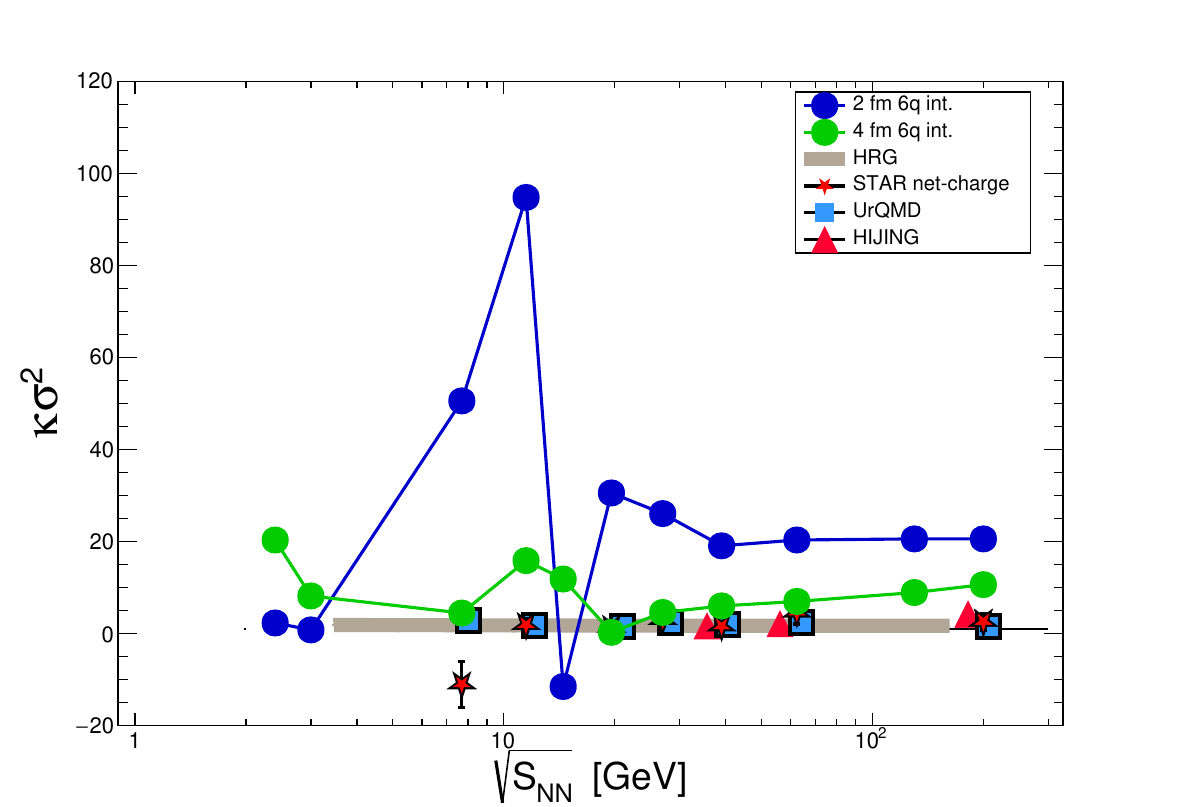} }}
    \caption{(color online) Beam-energy dependence of moment product $S\sigma^3/M$ (left) and $\kappa\sigma^2$($\frac{C_{4}}{C_{2}}$) (right) in the PNJL model with 6 quark interactions for finite volume systems with radius R=2fm and 4fm. For the system with the 6q-PNJL model, the blue color curve represents 2fm and the green color curve represents 4fm results. The shaded region (azure) in left side plot is from the Lattice QCD calculation with temperature T = 150-190 MeV. Both $S\sigma^3/M$ and $\kappa\sigma^2$ results with a function of center of mass energy are compared with the net-charge data of STAR experiments in red with black star. The results for $\kappa\sigma^2$ are compared with HRG, UrQMD and HIJING model calculations in grey, azure with black square and pink respectively}
    \label{fig:3}
\end{figure*}

The moment products $S\sigma^3/M$ (left) and $\kappa\sigma^2$($\frac{C_{4}}{C_{2}}$) (right) in the PNJL model with 6 quark interactions for finite volume systems with radius R=2fm and 4fm with respect to different center of mass energies are shown in figure 3. 

The values of \( S\sigma^3/M \) are compared with lattice QCD and STAR experimental data. The STAR results for net-charge fluctuations, obtained with \( 0.2 < P_T < 2 \) GeV/\(c\), \( |\eta| < 0.5 \), and 0--10\% centrality, along with lattice data corresponding to the temperature range 150--190 MeV, have been taken from Refs.~\cite{56,57} and included in the figure for comparison. The present 2~fm results show reasonable agreement with both the experimental and lattice QCD data in the energy range of 7--20~GeV, exhibiting visible fluctuations within this region. However, at higher energies, the 2~fm data tend to underestimate both the lattice and STAR results. The 4~fm results consistently underestimate the data in this energy range and remain largely energy-independent at higher energies.

The \( \kappa\sigma^2 = C_4/C_2 \) results, shown in Fig.~\ref{fig:3} (right), exhibit trends opposite to those observed in the \( S\sigma \) plot. The 2~fm 6-quark result shows larger fluctuations than the 4~fm 6-quark result up to 39~GeV. Beyond this energy, both curves appear nearly energy-independent. While minor fluctuations are present in the 4~fm result at lower energies, the trend becomes flat at higher energies.

These findings are compared with STAR experimental data on net-charge higher-order moments from RHIC BES-I (red with black borders), as well as with model predictions from HRG, UrQMD, and HIJING. Interestingly, the \( \kappa\sigma^2 \) values for the 4~fm system are in closer agreement with experimental results than the 2~fm values. The 4~fm curve generally follows the trend observed in the STAR data, except at 200~GeV, which may be attributed to resonance production at higher energies. Moreover, the 4~fm results align well with HRG, UrQMD, and HIJING simulations in the 19--40~GeV energy range, showing a rising trend at higher energies. In contrast, the \( \kappa\sigma^2 \) values for the 2~fm volume display non-monotonic behavior at lower RHIC energies, particularly between 3 and 30~GeV in center-of-mass energy and becomes independent thereafter.

\section{\label{sec:level4}SUMMARY AND CONCLUSIONS}

This study presents a detailed investigation of higher-order cumulants, $C_1$, $C_2$, $C_3$, and $C_4$ (moments, $M$, $\sigma$, $S$, $\kappa$) of net-charge fluctuations and their volume-independent ratios, $C_1/C_2$, $C_3/C_2$, and $C_4/C_2$ (products, $M/\sigma^2$, $S\sigma$, $\kappa\sigma^2$) as functions of center-of-mass energy, using the three-flavor Polyakov loop extended Nambu–Jona-Lasinio (PNJL) model with six-quark interactions at finite volume and finite density. These moments, directly related to generalized susceptibilities, are sensitive to the correlation length ($\xi$) in the system and serve as potential probes for identifying the QCD critical end point (CEP). Near the CEP, the correlation length and susceptibilities diverge, leading to enhanced non-Gaussian fluctuations that manifest as non-monotonic behavior of moment products with beam energy. The analysis covers a range of beam energies similar to those in the RHIC Beam Energy Scan (BES) program. Calculations of cumulants ($C_1$–$C_4$) are performed for systems with different finite volumes (2~fm and 4~fm), allowing exploration of system-size effects. The moment products are systematically compared with experimental results from the STAR Collaboration and theoretical predictions from HRG, UrQMD, and HIJING models. The results show that the 2~fm system exhibits pronounced non-monotonic behavior and better agreement with STAR data at lower energies, suggesting sensitivity to critical fluctuations in smaller systems. Conversely, the 4~fm system shows smoother, monotonic trends and aligns more closely with experimental data at higher energies. This indicates that both system size and interaction strength play crucial roles in the emergence and suppression of fluctuation signals near the CEP.

Overall, the study underscores the importance of finite-size effects and quark interactions in understanding the QCD phase structure and provides valuable input for ongoing and future experimental efforts to locate the QCD critical point.

\begin{acknowledgments}
P.Deb would like to thank Women Scientist Scheme A (WOS-A) of the Department of Science and Technology (DST) funding with grant no SR/WOS-A/PM-10/2019 (GEN).
\end{acknowledgments}

\end{document}